\documentstyle[psfig]{article}
\begin{document}
\date{}
\title{Negative Specific Heat in Astronomy, Physics and Chemistry}
\author{D. Lynden-Bell} \maketitle
\centerline{Institute of Astronomy, University of Cambridge,  CB3 0HA} 
\centerline{and Clare College,}
\centerline{Senior Fellow visiting The Queen's University, Belfast. BT7 1NN}

\begin{abstract}
Starting from Antonov's discovery that there is no maximum to the
entropy of a gravitating system of point particles at fixed energy in
a spherical box if the density contrast between centre and edge
exceeds 709, we review progress in the understanding of gravitational
thermodynamics.

We pinpoint the error in the proof that all systems have positive
specific heat and say when it can occur.  We discuss the development
of the thermal runaway in both the gravothermal catastrophe and its
inverse.

The energy range over which microcanonical ensembles have  negative
heat capacity is replaced by a  first order phase transition in the
corresponding canonical ensembles.  We conjecture that {\em all} first
order phase transitions may be viewed as caused by negative heat
capacities of units within them.

We find such units in the theory of ionisation, chemical dissociation
and in the Van der Waals gas so these concepts are applicable outside
the realm of stars, star clusters and black holes.
\end{abstract}

\section{Introduction}

When I first used the concept of Negative Specific Heat\cite{1} (or
more correctly Negative Heat Capacity) to explain Antonov's remarkable
Gravothermal \break Catastrophe\cite{2} the Statistical Mechanics community
thought I was talking nonsense.  But all astronomers had known since
the late 1800s that adding energy to a star or a star cluster would
make it expand and cool down\cite{3}.  Astronomers enjoyed this
incongruous result, but had not seen it as a real paradox in
thermodynamics until Thirring\cite{4} emphasised that there is a
theorem that specific heats are positive.  The `proof' as given in
Schr\"odinger's beautiful book\cite{5} is so simple that it is most
surprising that it could ever be WRONG.

Consider a canonical ensemble in thermal equilibrium
$$<\! E\! > = \sum_i E_i \exp \left(-\beta E_i\right) \bigg / \sum_i \exp
\left( - \beta E_i \right) \eqno (1)$$
$$C_V = {d<\! E\! > \over dT} = - k\beta^2 \, {d<\! E\! > \over d \beta} = k \beta^2
< (E_i - <\! E\! >)^2 >$$
the final expression, which follows from an elementary evaluation of
 $d<\! E \! > /d \beta$ via (1), is clearly positive, Q.E.D.  However,
the Astronomers' argument is hardly more complicated and involves the
kinetic energy $\cal T$.

The Virial theorem for a steady state under a potential energy, $\cal
V$, that scales like $r^{-n}$ reads [for external (or edge) pressure
$p_e$ and volume $V = {4 \over 3} \pi r^3_e$],
$$2 {\cal T} + n {\cal V} = 3 p_e V \ . $$
For gravity $n=1$ and the total energy, $E$, is the sum of kinetic and
potential parts.  So for an isolated gravitational system $(p_e = 0)$
$$E = - {\cal T} < 0$$
but for particles in motion ${\cal T} = {3 \over 2} NkT$ so 
$${dE \over dT} = C_V = - {3 \over 2} Nk \ , $$
which is clearly negative!

Luckily Thirring\cite{4} saw how to resolve the paradox and I like to
think he did so in favour of the astronomers but let me give you the
historical development without first unravelling the paradox.

Antonov\cite{2} took $N$ particles in a spherical box of radius $r_e$
and released them with energy $E$.  The total mass was $M = Nm$.  To
find the most probable state he looked for a local maximum in the
Boltzmann Entropy
$$S = -k \int f \, {\rm ln} \, f d^6 \tau$$
subject to the constraints
$$N = \int f d^6 \tau$$
$$E = \int f {p^2 \over 2m} \, d^6 \tau - {1\over 2} \int \! \! \int
Gm^2 {f f' \over |{\bf r} - {\bf r}'|}\,  d^6 \tau d^6 \tau' \ . $$
The integrals are taken over 6 dimensional phase space, $f$ is the
distribution function and $f'$ stands for $f ({\bf r}', {\bf p}')$.
The density $\rho ({\bf r}) = \int m f d^3 p$ and \break $d^6 \tau = d^3
rd^3p$,  As others had done, Antonov showed that the maximising $f$
was a function of $\epsilon = {1 \over 2} p^2/m - m \psi$ where $\psi$
is the gravitational potential determined self-consistently and this
leads to $\psi$ obeying the equation for the isothermal gas sphere, as
tabulated by Chandrasekhar\cite{6}.  We are all used to the entropy
being very sharply peaked about the most probable state so no-one
prior to Antonov had bothered to check whether the stationary entropy
states were in fact maxima.  Antonov found there was indeed a local
maximum whenever the radius, $r_e$, of the bounding sphere was not too
large, and, as $r_e$ was increased, the density at the edge, $\rho
_e$, dropped compared with that at the centre, $\rho _c$, due to the
gravity.  However, when $\rho_e/\rho_c$ dropped below $1/709$, he
found the point of stationary entropy ceased to be a local maximum and
became a minimax.  Moreover he was able to demonstrate that there is
no global maximum to $S$ at any fixed $E$ so that even when maxima
exist they are only local maxima.

Like many original results Antonov's were at first hard to believe.
It was particularly strange to me that the trouble occurred not when
the sphere was too small, where the gravity would be large, but when
the sphere was too large.  Nevertheless the effect was due to gravity
as it disappeared when gravity was absent.  To discover what was
happening I developed with Wood the thermodynamic theory of
self-gravitating gas spheres\cite{1}.  In particular we calculated the
energy $E$ and the heat capacity $C_V$ of isothermal spheres and drew
the $p_e, V$ diagram for adiabats.  Figures 1, 2 \& 3.

Figure 1 shows that $(-E) r_e/(GM^2)$ never exceeds 0.335 for any
equilibrium gas sphere, whatever the central concentration
$\rho_c/\rho_e$; so, if a mass $M$ of particles is released with
energy $E$ (negative) within a sphere of radius $r_e$ greater than
$r_A = 0.335 GM^2/(-E)$, there is no equilibrium state for them to go
to!  Furthermore when $r_e = r_A$ the density contrast is Antonov's 709.

Figure 2 plots $C_V = dE/dT$ as a function of the density contrast.
$C_V$ starts at the classical value $+{3 \over 2}Nk$ for a free gas
but increases and becomes infinite as the density contrast $\rho_c /
\rho_e$ approaches 32.2.  For larger density contrasts $C_V$ starts
negatively infinite and increases becoming zero at the Antonov point
where the density contrast is 709.  At this point the entropy ceases
to be a maximum at constant $r_e$ and $E$, so stability is lost and
the dashed prolongation of the sequence is unstable (see
Katz\cite{7,8} for the proof that they remain so).  However, isothermal
spheres with density contrasts between 32.2 and 709 are stable at
fixed $E$ and $r_e$ and have negative $C_V$.

Years of experience with positive $C_V$ systems are not a good
preparation for this, so let me review some obvious properties of
negative $C_V$ systems.

\begin{enumerate}
\item{Two negative $C_V$ systems in thermal contact do not attain
thermal equilibrium -- one gets hotter and hotter by losing energy,
the other gets for ever colder by gaining energy.  Thus negative $C_V$
systems can not be divided into independent parts each with negative
$C_V$; so negative $C_V$ systems are NEVER extensive.}
\item{A negative $C_V$ system can not achieve thermal equilibrium with
a large heat bath.  Any fluctuation that, e.g., makes it temporary
energy too high will make its temporary temperature too low and the
heat flow into it will drive it to ever lower temperatures and higher
energies.} 
\item{A negative $C_V$ system can achieve a stable equilibrium in
contact with a positive $C_V$ system provided that their combined heat
capacity is negative.  To see this imagine that the negative $C_V$
system `Minus' is initially a little hotter (higher $T$) than the
positive $C_V$ system `Plus'.  Then  heat will flow from Minus to
Plus.  On losing heat Minus will get hotter (i.e., its temperature
increases) but on gaining that heat Plus will also get hotter.
However, because Plus has a lesser $|C_V|$ its temperature is more
responsive to heat gain than Minus's is to heat loss.  Thus Plus will
gain temperature faster than Minus and a thermal equilibrium will be
attained with both Plus and Minus hotter than they were to start
with.  They also attain equilibrium if Minus is initially a little
cooler but the reader should think that through.

Notice that this stability is lost as soon as Plus has the same $|C_V|$
as Minus; i.e., when their combined heat capacity reaches zero from
below.
}
\end{enumerate}

Now we are in a position to explain Antonov's result via a thought
experiment.  Imagine a gravitating isothermal gas confined by a sphere
of radius just less than $r_A$ and adiabatically expand the sphere.
Work is done by the gas so $E$ becomes yet more negative and $r_A$
contracts while the sphere's radius expands making $r_e>r_A$.  The
inner parts of the gravitating gas are much denser and are held in
primarily by gravity so the expansion is mainly taken up by the less
dense gas in the outer parts.  Thus the adiabatic fall in temperature
of the outer parts due to their expansion will initially be greater
than the temperature fall of the inner parts.  Thus there will now be
a temperature gradient with the outer parts cooler than the central
ones.  However, as we have seen, the isothermal spheres with density
contrasts greater than 32.2 have negative specific heats.  As heat
flows down the temperature gradient the central parts contract and get
hotter while the outer parts held in by the sphere behave like a
normal gas so they receive the heat and get hotter too.  It is now a
race; do the outer parts get hotter faster on gaining the heat than
the inner parts do on losing it?  Clearly if the outer parts have too
great a positive heat capacity they will not respond enough and the
inner parts will run away to ever higher temperatures losing more and
more heat to the sluggishly responding outer parts.  This is Antonov's
gravothermal catastrophe.  Our criterion for this to happen is that
the combined heat capacities of the inner and outer parts should reach
zero from below and Figure 2 shows this happens at precisely Antonov's
point.  Thus isothermal gas spheres with density contrasts greater
than 709 are internally unstable and will spontaneously develop
temperature differences between centre and edge.  Katz \&
Okamoto\cite{9} recently emphasised that the effect of fluctuations
will decrease the 709 limit for simulations without very large $N$.

It is of interest to consider what happens at the end of the
gravothermal runaway as the centre gets hotter and hotter and denser
and denser.  If Antonov's point particles are replaced by small hard
spheres these will eventually get so close that they touch.  Then the
modified system will eventually achieve a new higher temperature
equilibrium centred on this condensed core.  Thus we suggested\cite{1}
that ``When the system approaches the gravothermal catastrophe the
system undergoes a phase transition in which a core of hard spheres in
contact with one another is formed''.  Aaronson \& Hansen\cite{10}
later showed this was so.  The more astronomically important case of a
non-relativistically degenerate Fermi Gas was likewise predicted to
show the same behaviour and give white dwarf configurations with total
masses below the Chandrasekhar Limit\cite{6}.  For higher mass systems
the Fermi Gas becomes relativistically degenerate with a
pressure-density relationship that is too soft to resist gravity, so
beyond the Chandrasekhar Limit the Gravothermal Catastrophe leads to
Black Holes.  The fact that these too have negative heat capacities
was demonstrated by Beckenstein\cite{11} and confirmed by
Hawking's\cite{12} exact calculation.  Earlier (1969) in contemplating
how quasars lived and died I predicted (Lynden-Bell\cite{13}) that
giant black holes, dead quasars, lay at the centres of most large
galaxies.  Although Maarten Schmidt early\cite{14} thought this was
quite likely, the idea gained acceptance only gradually until twenty
five years later Miyoshi et al.\cite{15} 1995 found a definitive case
in NGC4258.  Now Hubble Telescope results are widely interpreted as
showing giant black holes in many systems.

These are {\em not} the result of the gravothermal catastrophe of
stellar dynamics (see Section 3).  More energy loss via dissipative
gas dynamics and radiation from quasars is needed to get such large
masses into black holes by astrophysical processes.

Thirring's resolution of the paradox\cite{4} we posed in \S 1 is that
two negative specific heat systems can not be in thermal equilibrium,
so an equilibrium canonical ensemble of them is impossible.  Thus the
supposed `proof' that specific heats are positive implicitly assumes
the result and in reality only shows that {\em extensive} systems have
positive $C_V$.  

With Hertel\cite{16}, Thirring gave an example of a simple system with
a negative $C_V$ for a range of energies when considered in the
microcanonical ensemble.  When such systems were put in thermal
contact the whole showed a positive $C_V$ and underwent a
phase-transition that corresponded to the region of negative $C_V$ in
the microcanonical ensemble.

Some years later (1977) after Van Kampen challenged me over negative
$C_V$, my wife and I\cite{17} developed an easily calculable
gravitational model which demonstrated these effects and predicted the
temperature of the gravitational phase transition in the corresponding
canonical ensemble.  The model has many equal particles confined on a
sphere of variable radius $r$ and is governed by the Lagrangian
$$L = \sum^N_{i=1} {\textstyle {1 \over 2}} mr^2 \left( {\dot
\theta}^2_i + \sin^2 \theta {\dot \phi}^2_i \right) + {\textstyle {1
\over 2}} M {\dot r} ^2 + {\textstyle {1 \over 2}} GM^2/r, \ \ r_0
\leq r \leq r_e$$
and $r$ is constrained to lie between some small $r_0$ and some large
$r_e$.  The particles on the sphere share energy as in a perfect gas.
This system has $C_V = - (N + {\textstyle {1 \over 2}}) k$ whenever $a
\equiv {\textstyle {1 \over 4}} GM^2/(-E)$ lies between $r_0$ and
$r_e$.  However, it has positive $C_V$ whenever $r$ is up against one
of its limits, i.e., whenever `$a$' lies outside the range $(r_0, r_e)$,
which of course happens at very low or very high energies, see Figure
4.  When we considered many of these whole systems in a canonical
ensemble the individual spheres were found against one stop or the
other even when the mean energy was well within the microcanonical
region of negative $C_V$.  The ensemble now had $C_V >0$ and a
macroscopic phase transition with a latent heat corresponding to the
energy difference between the spheres on one stop and those on the
other.  Adding energy (heat) merely led to some of the systems
formerly against the lower stop, $r_0$, migrating around the curve to
the higher stop.  Thus even macroscopic systems that have negative
$C_V$  over a wide range of energies behave quite differently when
thermally coupled {\em at equilibrium}.  The whole assembly has a
positive $C_V$ and a phase transition which we determined in detail,
see Figure 4.  This behaviour led us to ask whether {\em all} first
order phase transitions could be viewed as due to negative specific
heat systems of molecular size that might cause all phase transitions
as discussed in the next section.  

However, before leaving this one it
is important to discuss what happens in practice as well as what
happens `at equilibrium' especially as the metastable superheated or
supercooled regions in our model can be large.  These metastable
regions have the sphere against one stop or the other.  Even in the
canonical ensemble it takes a gigantic fluctuation to take such a
metastable system over the hump.  It has to gain a thermal energy
comparable to the mean energy of each whole system, i.e., it needs a
fluctuation $\sqrt{N}$ times the typical one.  As emphasised by
Parentani, Katz \& Okamoto\cite{18} for the black hole case, such
fluctuations will occur almost never -- i.e., once in $\exp (-N)$
independent trials.  Thus to get the thermodynamic phase transition
$N$ can not be large and even for $N=10$ one must wait for $\sim
e^{10}$ complexions of the system.  Thus these metastable regions of
our large systems will be exceedingly stable and no phase transition
of the canonical ensemble will be observed until the system nears the
top of Figure 4a.  However, once on the unstable branch the system
will evolve along the downslope on the timescale of thermal diffusion
and will then become so much cooler than the ensemble that on a
similar timescale it climbs the other branch to regain the `tops'
temperature of $(N+ {\textstyle {1 \over 2}})kT \sim {\textstyle {1
\over 2}} \chi_0$.  If the ensemble is cooled, likewise the transition
will not be observed until the temperature reaches the much lower
temperature of $(N+ {\textstyle {1 \over 2}})kT \sim {\textstyle {1
\over 2}} \chi_e$ when the transition will again go violently.  Thus
there will in practice be an enormous hysteresis compared to which
`boiling with bumping' will look quite petty. 

Kiessling\cite{19} has emphasised that statistical mechanics done
properly gives transitions like $AC$ rather than ones associated with
the ends of the metastable regions but for transitions in these
macroscopic systems the path $AC$ would not be encountered in practice
nor normally in simulations unless the numbers of particles involved
are small.  However, the metastable regions in microscopic systems can
be surmounted by fluctuations with smaller energy changes so ice melts
at the temperature at which water freezes rather than at a
significantly higher one.

\section{A Different View of the Cause of Phase Transitions}

In the last section we saw how a region of $C_V <0$ in individual
systems gave rise to a phase transition when many such systems were
placed in thermal contact.

In reference 13 we asked whether {\em all} phase transitions (or at
least all first order ones) can be viewed as caused by negative
specific heat elements at a molecular level.  Can one look at systems
that undergo changes of state and identify such negative-heat-capacity
elements?  We shall demonstrate their existence for simple models of
chemical dissociation or ionisation equilibrium and in systems like
the Van der Waals gas.

We consider two types of particles, $A$ and $B$ which are in the
ground state tied together in pairs $AB$.  We take the forces between
them to saturate when such a pair exists and shall model the
interaction potential to be a $\delta$ function so a single bound
state exists with binding energy $\chi$.  Let $n_{AB}$ be the number
density of pairs in some initial ground state.  We shall model this
system by a microcanonical ensemble of boxes each of volume $L^3 =
n^{-1}_{AB}$ and each containing one particle of type $A$ and one of
type $B$.  We shall show that each of these microsystems shows a
negative specific heat when the energy of each box is just greater
than the dissociation energy.  Taking that as our zero point for
energy we find the number of energy levels $<E$ for a system of two
independent free particles is
$$\tau_f (E) \propto \left\{ {E^3L^3 \atop 0} \ , \ {E \geq 0 \atop
E<0} \right. \ , $$
while for a bound pair the number of energy levels is $\tau _b (E)
\propto \left[ (E + \chi) L^2 \right]^{3/2}$.  

The total number of energy levels $<E$ is thus $\tau (E) = \tau_f +
\tau _b$.  Gibbs gives both $S_1 = k \, {\rm ln} \, \left( d \tau / d
E \right)$ and $S_2 = k\, {\rm ln} \, (\tau)$ as possible expressions
for the entropy for small systems.  We earlier\cite{17} showed $S_1$
gave negative specific heat just above the dissociation point so here
we consider $S_2$.  Evidently
$$kT = k {dE \over dS_2} = {\tau \over \tau '} = {(E + \chi)^{3/2} +
AE^3 \over {3 \over 2} \left[ (E + \chi )^{1/2} + 2 AE^2 \right]} \ ,
$$
where $A \propto L^{3/2}$.  It is not hard to show that $${k \over C_V} =
{dkT \over dE} = \left( {\tau \over \tau '} \right)'$$
is negative for some (large enough) values of $L$ for energies greater
than zero and on up to about $0.8 \chi$ for some $L$.  Thus if the
system is in a large enough box, i.e., rare enough, too many
dissociate as the energy is increased so the mean kinetic energy per
free particle {\em decreases} because of the energy soaked up by the
dissociation.  In this sense negative $C_V$ elements are associated
with chemical dissociation reactions.  Much the same arguments hold
for ionisation reactions at higher temperatures.

Perhaps of greater interest is the demonstration of similar phenomena
in the Van der Waals gas and the following demonstration is due to
\break R.M. Lynden-Bell my wife.  We are all familiar with the dip and
hump in the isotherms of the Van der Waals gas and how Maxwell's
construction replaces them with the constant pressure phase-transition
that occurs in practice.  This is what will happen in an extensive
system with very many identical subsystems.  However, at or close to
the molecular level the extensivity breaks down because such tiny
systems can not be readily divided into two equivalent pieces.  The
Van der Waals equation includes such effects in its molecular volume
term $b$ and its mean molecular attraction term which reduces the
effective pressure by $a \rho^2$.  We shall assume that at a molecular
level the tiny {\em indivisible elements of Van der Waals gas actually
obey Van der Waals's equation} and it is only a cooperative effect of
many of them in a canonical ensemble that makes the ensemble obey the
Maxwell construction and give the phase transition.  We saw that this
is precisely what happened when we took our gravitating systems in
Section 1 and put many of them into a canonical ensemble at
equilibrium.  What we now show is that an element that obeys the full
curve of the Van der Waals equation inevitably has a negative $C_p$.
The hump in the Van der Waals isotherms at temperatures below the
critical one gives two stable states $AB$ which share the same
temperature and pressure (the third is unstable).  Thus
$$0 = \Delta p = \int^B_A \left( {\partial p \over \partial V} \right)_T dV \ . $$
For this to happen $(\partial p / \partial V)_T$ can not have the same
sign everywhere so we need an unusual region with it positive.

Now $$C_p^{-1} = \left( {\partial ({\rm ln}T) \over \partial S}
\right)_p$$ so normally, with $C_p$ positive, ${\rm ln}T$ will increase
with $S$ at constant $p$.  However, the states $A$ and $B$ share a
common isotherm and are at the same pressure so
$$0 = \Delta ({\rm ln}T) = \int^B_A \left( {\partial {\rm ln}T \over
\partial S} \right)_P dS = \int^B_A {1 \over C_p} dS \ , $$
so $C^{-1}_p = \left( \partial {\rm ln} T/\partial S \right)_p$ must
reverse sign and the graph of ${\rm ln}T$ against $S$ can not be
monotonic.  We have, therefore, demonstrated that for any system which
has a hump in its isotherm like that in the Van der Waals gas there is
a region of negative $C_p$.  This is our prime result!

One learns at school that for perfect gases $C_p - C_V = R$ so one
might expect that systems with $C_p$ negative would also have $C_V$
negative.  This is not so,
$$\displaylines{C_p - C_V = T \left( {\partial S \over \partial T}
\right)_p - T \left( {\partial S \over \partial T} \right)_V = T
\left( {\partial S \over \partial V} \right)_T \left( {\partial V
\over \partial T} \right)_p = \hfill \cr \hfill{} = T \left( {\partial
p \over \partial T} \right)_V \left ( {\partial V \over \partial T}
\right)_p = - \left( {\partial p \over \partial V} \right)_T T \left
( {\partial V \over \partial T} \right)_p^2 \ .} $$ Thus $C_p$ is only
greater than $C_V$ at positive $T$ when, as is usually the case,
$\left( {\partial p \over \partial V} \right)_T$ is negative.  But
when the pressure along an isotherm increases with $V$, as in a phase
transition region, $C_p$ is actually less than $C_V$.  It is,
therefore, a moot point whether $C_V$ has to be negative.  Here Van
der Waals's gas is of great interest as an example because for it
gives $C_V = {3 \over 2}Nk$ which is positive everywhere!  Thus the
hump and trough in the isotherm is associated with negative $C_p$ but
not with negative $C_V$.

\section{Aftermath of the Catastrophe}

We earlier left our gas just undergoing the gravothermal catastrophe
with the central part contracting and getting hotter while the outer
part could not raise its temperature fast enough to keep up.  Since
the centre is now denser the Antonov point at which the density
reaches $\rho_c/709$ is now inside the system rather than at its
edge.  Thus this point moves inwards through the mass.  In astronomy
we are primarily interested in applying this theory to a `gas' in
which each molecule is replaced by a star so that we deal with a star
cluster or the nucleus of a galaxy.  In such systems the timescale for
exchanging the `heat' of random stellar motion is shortest at the
centre and can become very long in the outer parts.  As the Antonov
point moves in, the timescale of heat flow becomes shorter and the
regions beyond the Antonov point are too sluggish to respond other
than adiabatically.  The gravothermal catastrophe occurs again and
again nay continually, with ever higher densities and temperatures at
ever smaller scales\cite{20, 21, 22} and the precise form of the 
initial conditions is rapidly forgotten.  Since the same process is
occurring at ever smaller scales we expect a similarity solution with
the density of the form
$$\rho (r,t) = \rho _c (t) \rho_\star (r_\star)$$
where $r_\star = r/r_c (t)$ and $r_c(t)$ the core radius will be some
fixed fraction of the Antonov radius $r_A$ where the density is
$\rho_c/709$.  Since the halo is so sluggish that it is left behind in
the every quickening evolution of the centre we can put $\partial
\rho/\partial t = 0$ at large $r$.  Hence $\dot \rho _c \rho_\star -
{\dot r_c \over r_c} r_\star \rho_c \rho'_\star = 0$ and so we may
separate the $t$ and $r_\star$ dependences to get
$${r_\star \rho'_\star \over \rho_\star} = {r_c \dot \rho _c \over
\dot r _c \rho _c} = - \alpha \ . \eqno {\rm (large} \ r)$$
Since the $\star$ variables are independent of the $t$ variables
$\alpha$ must be a constant.  Thus $\rho _\star = A r_\star^{-\alpha}$
at large $r_\star$ and $\rho_c \propto r^{-\alpha}_c$ for all $t$.

Now the evolution of the system is due to the heat transport whose
rate is determined by ${\dot \rho_ c \over \rho _c} \propto {1 \over
T_r}$ where $T_r$ is the relaxation time within the Antonov radius;
for it the standard formula is $T^{-1}_r \propto {\rho _c \over v^3_c}
(8 \pi G^2m \, {\rm ln}\, N)$ where $v^2_c$ is the velocity dispersion
of the stars. In our self-similar collapse $$v^2_c \propto G {4 \over 3}
\pi \rho_c r^3_c/r_c \propto G \rho_c r^2_c \propto G
\rho^{1-2/\alpha}_c\ . $$  Hence $${\dot \rho _c \over \rho _c} \propto
\rho_c^{{3 \over \alpha} - {1 \over 2}} \ \ .$$On integration we
find\cite{21,23} $$\rho_c \propto |t_0 - t|^{-2 \alpha / (6 - \alpha)}
\ \ \ {\rm so} \ \ \ r_c \propto |t_0 - t |^{2/(6 - \alpha)} \ \ . $$
Thus the core radius becomes zero and the central density formally
$\rightarrow \infty$ at $t_0$.  The core mass is $$M_c \propto \rho _c
r^3_c \propto |t_0 - t|^{2(3 - \alpha)/(6 - \alpha)}$$ and the core
velocity dispersion behaves as $$v_c^2 \propto GM_c/r_c
 \propto |t_0 - t|^{(4 -2 \alpha)/(6 - \alpha)}\ . $$We must still
determine $\alpha$.

Now at large distances $\rho_\star \propto r_\star^{- \alpha}$  and
the asymptotic form of the constant temperature (isothermal) sphere
is $\rho \propto r^{-2}$ so for an outward temperature decrease
$\alpha >2$.  However, if $\rho \propto r^{-2.5}$ there is an infinite
binding energy near the centre and to create this in finite time would
need an impossibly large heat flux.  Thus from these general arguments
$2 < \alpha < 2.5$.  Detailed calculations give $\alpha$ as an
eigenvalue and for stellar dynamics all investigators get \break $\alpha =
2.22 \pm .01$ which gives the rather weak dependence $v_c \propto
|t_0 - t|^{-(\alpha - 2)/(6 - \alpha)}$.  From these one finds $M_c
\propto v_c ^{-2(3-\alpha)/(\alpha - 2)}$ where the exponent is 7.1
for $\alpha = 2.22$.  If in core collapse $v_c$ increased from 300
km/s to 300,000 km/s corresponding to a black hole then $M_c$ would
decrease by a factor $2 \times 10^{21}$ leaving much less than one
star so these conditions are not attained.  In fact if $M_c$ decreased
by $10^{7.1}$ then $v_c$ would increase by a factor 10 only.  However,
in fact a new phenomenon sets in a high star densities that produces a
delightful new twist in the evolution.  Henon\cite{24} in his most
percipient early work suggested that the formation of binary stars
would occur at very high densities and that this would produce a new
energy source.  Heggie\cite{25} did the seminal work on the binary creation
rate and Betteweiser \& Sugimoto\cite{26} put this into their code for core
collapse.  As Sugimoto had surmised\cite{27}, binary formation sets in quite
suddenly during core collapse and releases energy near the centre.
Since the core is of negative heat capacity it immediately expands and
becomes of lower temperature than its surroundings.  The stage is now
set for the inverse gravothermal catastrophe.  As heat is conducted
into the core it grows and gets colder.  The process goes on until
eventually the core grows so large that it contacts the cooler outer
parts of the halo.  Then the whole core becomes isothermal and
eventually feels the heat loss to the outside so the gravothermal
collapse stars again.  A number of these giant thermal pulses occur in
simulations of post-collapse core evolution so it is hard to tell
whether the self-similar evolution\cite{23} predicted on the basis of
a singular core from which a flux of energy emerges due to continued
energy emission from binaries gives even a roughly correct average
evolution.  

Padmanabhan\cite{28} has given a longer more quantitative review of
gravitational thermodynamics which discussed many of the phenomena
considered here.

Meylan \& Heggie\cite{29} have an up to date review of the more
astronomical aspects.  

Miller \& Youngkins\cite{30,31} have given the mean field theory for a
simple system of spheres and have conducted numerical simulations to
verify the existence of gravitational phase transitions.  

Both Miller \& collaborators\cite{32} and Tsuchiya\cite{33} have
conducted very long term integrations to investigate the approach to
equilibrium of a one-dimensional system of mass sheets.  For these
relaxation is particularly slow as the interactions when they cross
are rather smooth.

\newpage

\begin{figure}[th]
\centerline{\psfig{figure=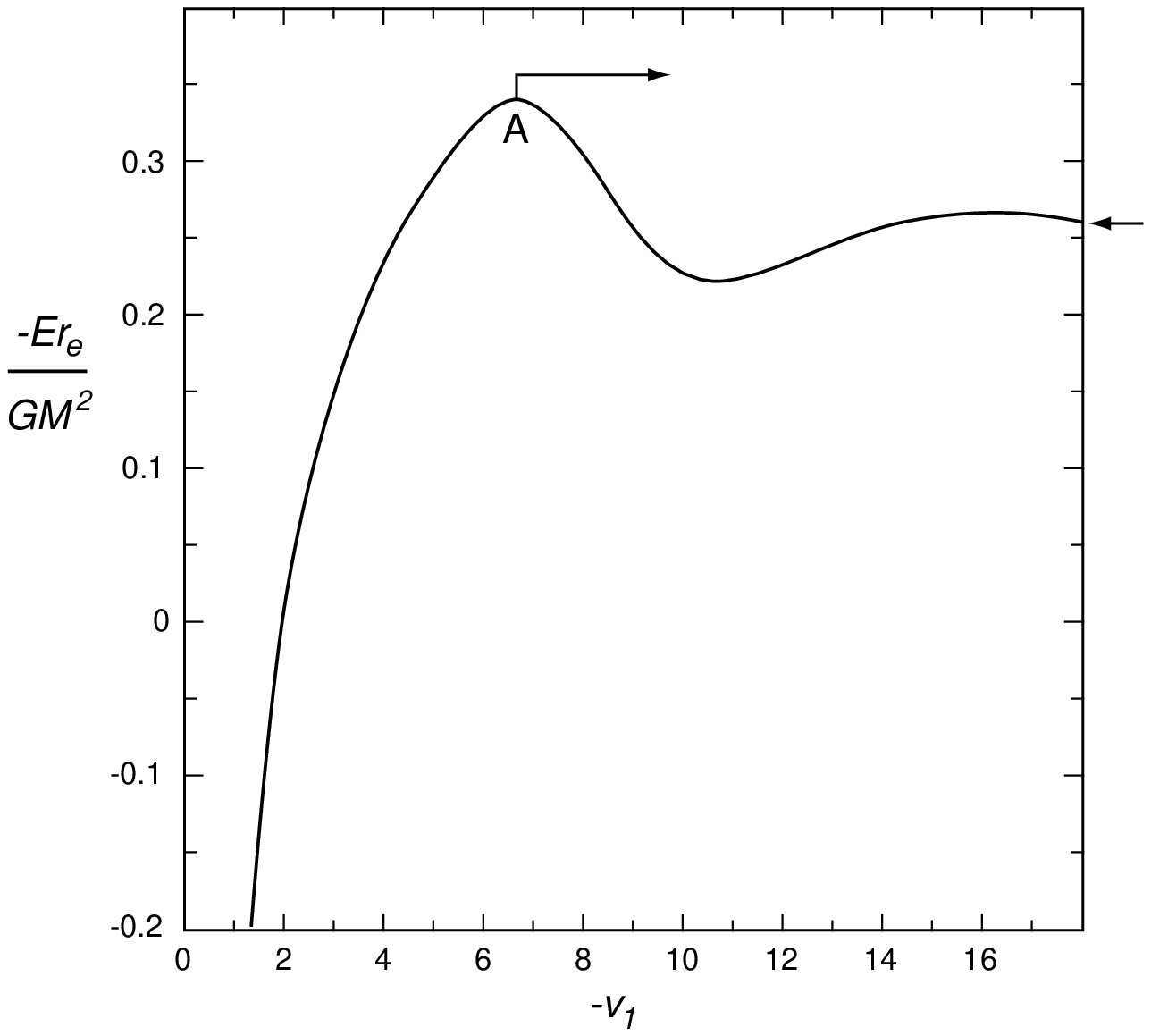,height=4in}}
\caption{The dimensionless binding energy $-Er_e/(GM^2)$ of an
isothermal gravitating sphere of mass $M$ in a spherical container of
radius $r_e$ plotted as a function of the density contrast, ${\rm
ln}\rho_c/\rho_e = -v_1$, between centre and edge.  Instability sets in at
$A$, the maximum.}
\end{figure}

\newpage

\begin{figure}[th]
\centerline{\psfig{figure=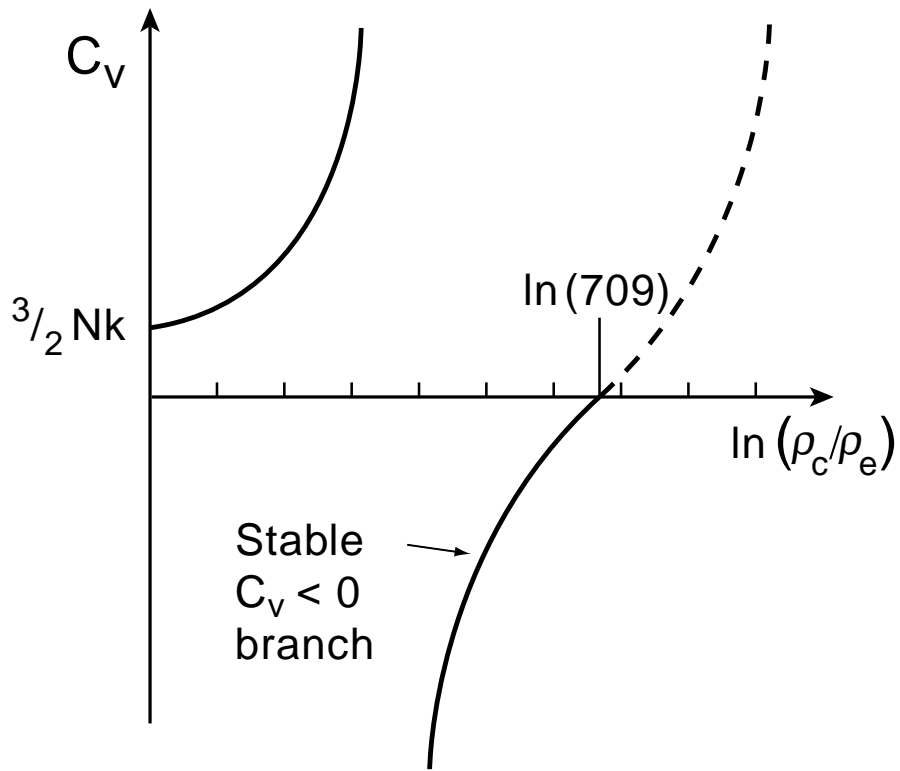,height=4in}}
\caption{The specific heat $C_V = dE/dT$ for a self-gravitating
sphere in a spherical container plotted as a function of the density
contrast.  Instability sets in at $A$ as $C_V$ reaches zero FROM
BELOW!}
\end{figure}

\newpage

\begin{figure}[th]
\centerline{\psfig{figure=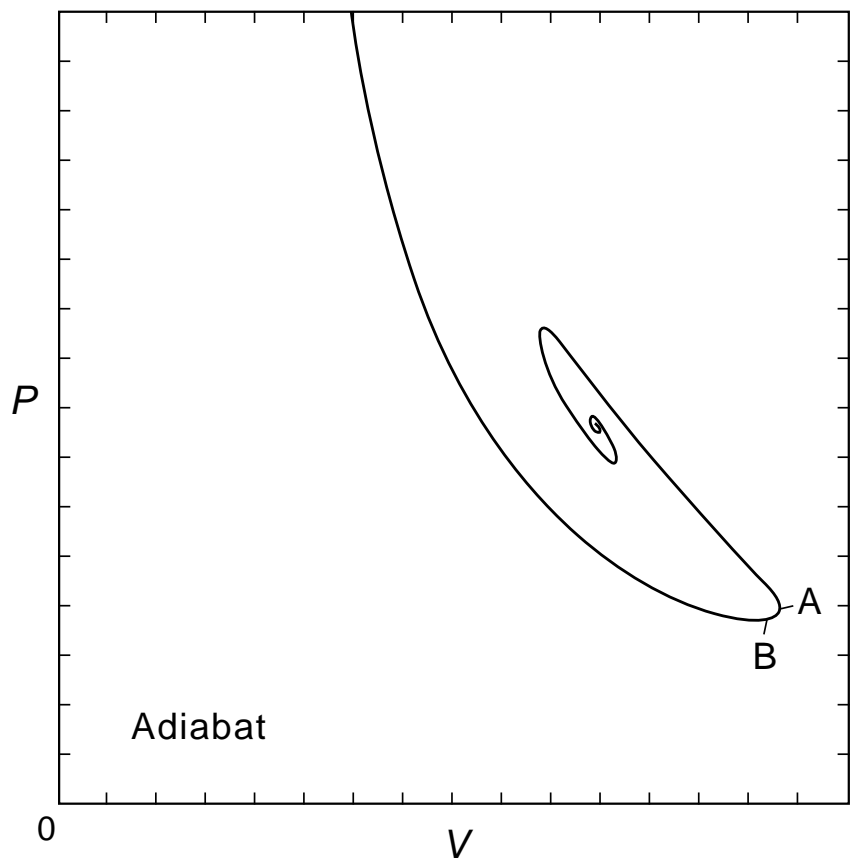,height=4.5in}}
\caption{The adiabat plots the surface pressure of the sphere against
its volume $V = {4 \over 3} \pi r^3_e$ at constant entropy,
instability sets in at $A$ for an isolated system but at $B$ if the
system is held at constant pressure.}
\end{figure}

\newpage

\begin{figure}[th]
\centerline{\psfig{figure=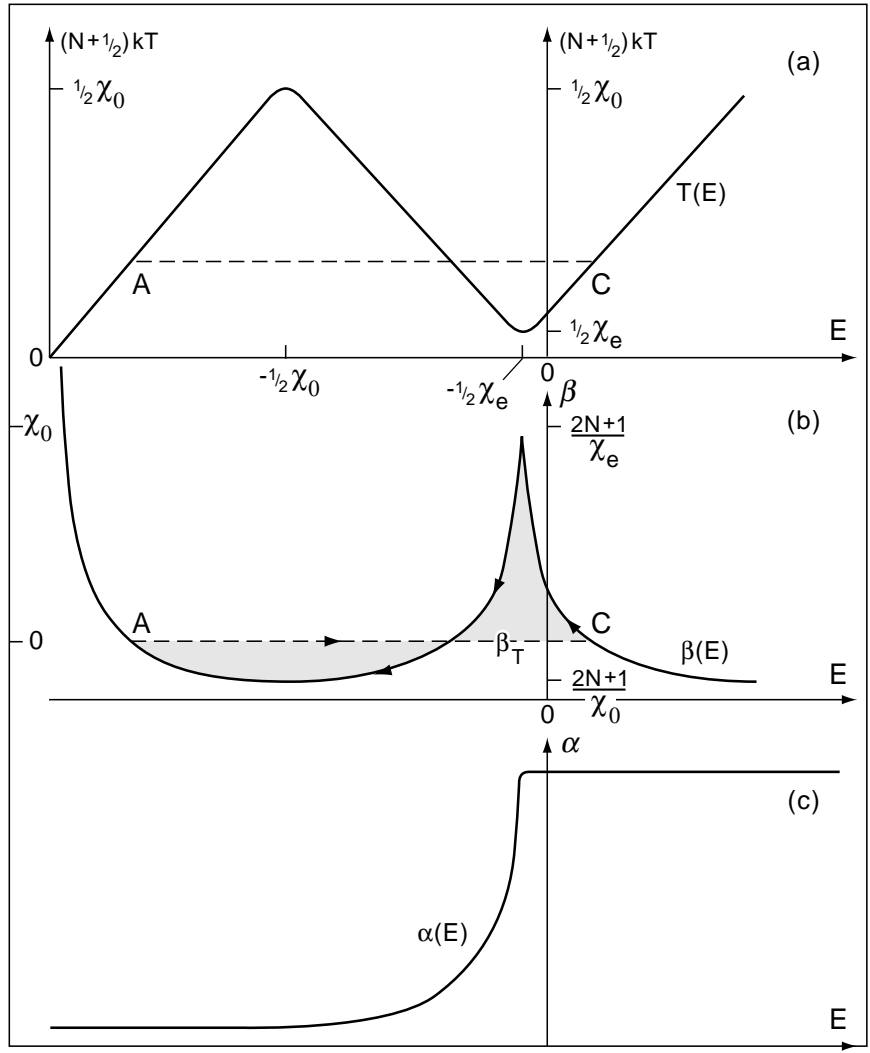,height=5.5in}}
\caption{(a) $T$ as a function of $E$ for the variable radius
slippery sphere model.  $AC$ is the phase transition predicted for a
canonical ensemble but replaced in practice, if $N$ is large, by lines
to the maximum ($T$ increasing) or minimum ($T$ decreasing) of the
$T(E)$ curve which then shows a hysteresis loop. (b)
$(kT)^{-1}$ as a function of $E$.  The shaded regions are of equal
area for the canonical phase transition. (c) The mean radius
of the sphere as a function of energy.}
\end{figure}

\end{document}